\documentclass[fleqn,twoside]{article}
\usepackage{epsf,multicol,ifthen}
\usepackage{ujp}
\usepackage[cp1251]{inputenc}
\usepackage[english,russian]{babel}
\usepackage{amstext}
\usepackage{amssymb}
\usepackage{amsmath}
\usepackage{cite}
\usepackage {graphicx}

\makeatletter
\renewcommand{\thesection}{\arabic{section}}
\renewcommand{\p@subsection}{}

\renewcommand{\p@subsubsection}{}

\makeatother
\newcommand*{\dis}{\displaystyle}
\newcommand*{\bs}{\boldsymbol}
\newcommand*{\intl}{\int\limits}
\newcommand*{\suml}{\sum\limits}

\nazva{THERMALIZATION IN HEAVY-ION COLLISIONS}%

\nazvacol{THERMALIZATION IN HEAVY-ION COLLISIONS}%

\avtor{D.V.~ANCHISHKIN, \, S.N.~YEZHOV$^{1}$}%
\avtorcol{D.V.~ANCHISHKIN, S.N.~YEZHOV}%

\inst{M.M.~Bogolyubov Institute for Theoretical Physics, National
Academy\\ of
Sciences of Ukraine}%
\adr{(14b, Metrologichna Str., Kyiv 03680, Ukraine; e-mail:
anch@bitp.kiev.ua), }
\insti{Taras Shevchenko Kyiv National University }%
\adri{(2, Academician Glushkov Ave., Kyiv 03022, Ukraine; e-mail:
yezhov@univ.kiev.ua)}%

\begin{document}
\setcounter{page}{1}
\maketitl
\begin{multicols}{2}
\anot{%
We propose a model for isotropization and corresponding
thermalization in a nucleon system created in the collision of two
nuclei.
The model is based on the assumption: during the fireball evolution,
two-particle elastic and inelastic collisions give rise to the
randomization of the
nucleon-momentum transfer which is driven by a proper distribution.
As a first approximation, we assume a homogeneous distribution where
the values of the momentum transfer is bounded from above.
These features have been shown to result in a smearing of the particle
momenta about their initial values and, as a consequence, in their
partial isotropization and thermalization.
The nonequilibrium single-particle distribution function and
single-particle spectrum which carry a memory about initial state
of nuclei have been obtained. }

\section{Introduction}

The problem of isotropization and thermalization in the course of
collisions between heavy relativistic ions attracts much attention,
because, while describing experimental data, the application of
thermodynamic models is one of the basic phenomenological
approaches. Recently \cite{kovchegov}, this issue was examined for
quark-gluon plasma produced as a result of ultra-relativistic $A+A$
collisions in experiments at CERN (Geneva) and BNL (Upton)
\cite{heinz}.

Nowadays, there exist a few interesting models for the explanation of the
thermalization phenomenon in parton systems. Among them, there is a model,
which considers, in the framework of quantum chromodynamics, the influence
of external color fields on vacuum with the following creation of
thermalized system of hadrons \cite{dokshitzer}. There is also a model of
thermalization in heavy ion collisions, which is considered as a consequence
of the Hawking--Unruh effect \cite{kharzeev}.

Such an enhanced attention to those phenomena arose, because, along
with other factors, the assumption about a local thermodynamic
equilibrium is successfully applied in various domains of
high-energy physics.
For instance, the transverse spectra of hadrons
look thermalized not only if heavy ions collide, when the creation
of many-particle statistical systems is adopted as an indisputable
fact \cite{shuryak}, but also at an $\mathrm{e}^{+}\mathrm{e}^{-}$
annihilation (see work \cite{becattini-hep-ph-0410403}), when the
issue concerning the creation of many-particle system remains open for
discussion.

In this work, we attempted to find an explanation for the
thermalization phenomenon, starting from such basic concepts as the
conservation laws of energy and momentum, and leaving details, which
are characteristic of every specific process, aside for further
investigation.
We offer a model, to a large extend simplified, which
we called \textit{the maximal isotropization model} (MIM). This
model belongs, to a certain degree, to transport ones; we consider
the evolution of the system, but we parametrize this development by
the number of collisions of every particle in the system rather than
by the time variable. The idea of maximal isotropization consists in
that, we suppose that a nucleon momentum transfer after every three
nucleon-nucleon (-hadron) collisions becomes a random quantity
driven by a proper distribution.

Let us assume for the moment that elastic scattering is the main
contribution to the two-nucleon collision amplitude.
Owing to
the features of collisions between heavy ions, the initial (before
the first collision) momenta of particles in $N$-particle system
$A$ and $N$-particle system $B$ are known exactly.
More precisely speaking, the initial momentum of every nucleon in the
nucleus $A$ is $\boldsymbol{k}_a=\boldsymbol{k}_0=(0,0,k_{0z})$,
while the initial momentum of every nucleon in the nucleus $B$ is
$\boldsymbol{k}_b=-\boldsymbol{k}_0=(0,0,-k_{0z})$.
The energy and momentum are conserved in every separate collision of
two particles
\begin{equation}
 \omega({\bs k}_a) + \omega({\bs k}_b) = \omega({\bs p}_a) +
\omega({\bs p}_b), ~~ \bs k_a + \bs k_b  =  \bs p_a + \bs p_b ,
\label{i1}
\end{equation}
where $\boldsymbol{k}_{a}$ and $\boldsymbol{k}_{b}$ are the initial
momenta of this particle pair, while $\boldsymbol{p}_{a}$ and
$\boldsymbol{p}_{b}$ are the corresponding final momenta.
We assume that the particles are on the mass shell, so that
$\omega (\boldsymbol{k})=\sqrt{m^{2}+\boldsymbol{k}^{2}}$.
We also adopt the system of units, where $\hbar =c=1$.

So, after the first collision, we have only four equations for the
determination of six unknown quantities, $\boldsymbol{p}_{a}$ and
$\boldsymbol{p}_{b}$.
It means that two quantities, e. g.,
$(\boldsymbol{p} _{a})_{x}$ and $(\boldsymbol{p}_{b})_{x}$, remain
uncertain and can be considered as such which accept random values,
which are driven by a scattering probability.
After the third collision, every component of the particle
momentum becomes completely random.
If the initial momentum is fixed this means a full randomization of
the momentum transfer after three serial collisions.
So, if we follow nucleon sequential elastic scattering from the first
collision to the last one we would see
a full randomization of the momentum transfer after every three
sequential scattering.
%
%
\begin{center}
  \begin{minipage}{0.9\columnwidth}
\includegraphics[width=1.0\textwidth]{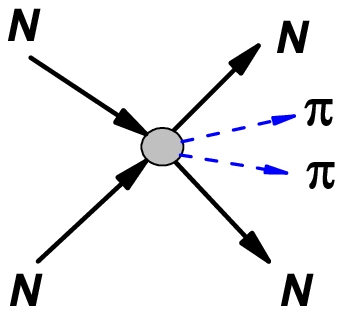}
{\footnotesize Fig.~1.~Example of inelastic collision of two
nucleons with creation of two $\pi$-mesons,
$N+N \to N+N+\pi + \pi$.
}
  \end{minipage}
\end{center}
%

In the inelastic collisions, the nucleon momentum transfer undergoes
even faster randomization.
Indeed, let us consider, for example, the process of creation of
$n_\pi$ pions in the nucleon-nucleon reaction:
$N+N \to N+N+n_\pi\cdot \pi$.
The process leads to randomization of $[3(2+n_\pi)-4]$ degrees of
freedom, or $d_p$ degrees of freedom per particle
becomes random, where $d_p=3-4/(2+n_\pi)$ and
we assume that all particles are on the mass-shell.
For instance, if $n_\pi=2$  (see the physical diagram of the collision
depicted in Fig.~1), then two components of the momentum of every
particle after reaction becomes random, i.e. $d_p=2$ in this process.
If $n_\pi \gg 1$, the number of the random degrees of freedom per
particle achieves its maximal value $d_p=3$.
We see that in the inelastic collisions the randomization of the
nucleon momentum transfer, attributed to one physical diagram, goes
faster than in the elastic scattering.

To estimate the effective number of collisions, $M$, we analyze all
nucleon collisions (physical diagrams), $N_{\rm coll}$, and obtain a
total sum of the
random degrees of freedom, $d_{\rm tot}$, gained by the particle
during all reactions before the freeze-out, i.e. we need to know
$d_{\rm tot}=\sum_{i=1}^{N_{\rm coll}} d_p^{(i)}$.
Then, the effective number of collisions is determined as
$M=d_{\rm tot}/3$.

So, in the framework of the MIM we assume a randomization of the
momentum transfer, $\bs p$ (see Fig.~2), after every effective
collision and the number of effective collisions, $M$, is determined
for a particular experiment.
At the same time, we assume that the value of the momentum transfer
is bounded from above by a maximal value $p_{\rm max}$.
Here, we based on the experimental fact that
%
%
\begin{center}
  \begin{minipage}{0.9\columnwidth}
\includegraphics[width=1.0\textwidth]{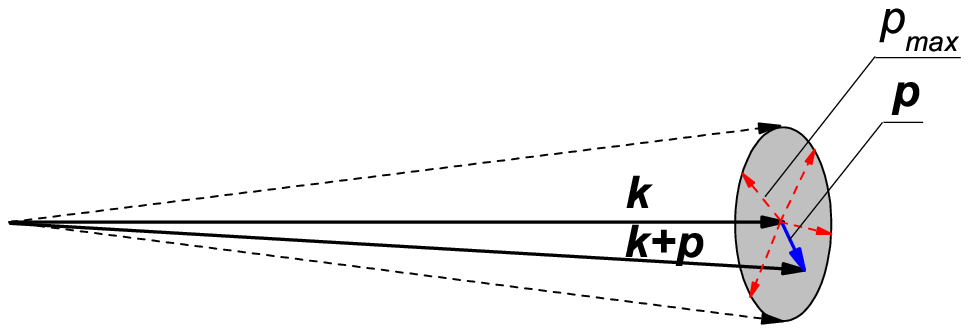}
{\footnotesize Fig.~2.~Nucleon momentum transformation as a result
of two-particle collision; $\bs k$ is the initial momentum,
$\bs p$ is the momentum transfer,
$\bs k + \bs p$ is the momentum after reaction,
$p_{\rm max}$ is the maximally allowed value of the momentum transfer.
}
  \end{minipage}
\end{center}
%
in hadron-hadron and
nucleus-nucleus collisions at high energies the big values of the
momentum transfer are suppressed and the eikonal approximation is
appropriate for description of two-particle scattering and
multi-particle production.

\section{Particle Distribution in Momentum Space}
As a building block for more complicate picture of relativistic
nucleus-nucleus collisions we consider an
idealized many-particle system where every particle experiences the
identical number of effective collisions, $M$, before system decay.
Indeed, finite number of collisions is a common feature of the system
with finite life time.
Consider successive variations of the momentum of the $n$-th nucleon
from nucleus $A$, which moves along the collision axis from left to
right.
Every $m$-th collision induces the momentum transfer,
${\boldsymbol{p}} _n^{(m)}$, for the $n$-th nucleon so that, after
$M$ collisions, the nucleon acquires the momentum
${\boldsymbol{k}}_n$:
\begin{equation*}
{\bs k}_{0} \rightarrow  {\bs k}_{0}+{\bs p}_{n}^{(1)}
\rightarrow  {\bs k}_{0}+{\bs p}_{n}^{(1)}+{\bs p}_{n}^{(2)}
\rightarrow \cdots
\rightarrow {\bs k}_{0}+{\bs P}_n\equiv
{\bs k}_{n}\,,
\end{equation*}
where ${\bs P}_n=\sum_{m=1}^{M}{\bs p}_{n}^{(m)}$, see Fig.~3.

%
\begin{center}
  \begin{minipage}{0.9\columnwidth}
\includegraphics[width=1.0\textwidth]{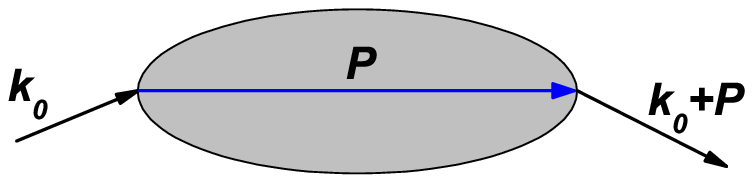}
{\footnotesize Fig.~3.~Transformation of the initial nucleon momentum,
$\bs k_0$, as a result of $M$ collisions;
${\bs P}=\sum_{m=1}^{M}{\bs p}^{(m)}$ is the total
momentum transfer after $M$ effective collisions.
 }
  \end{minipage}
\end{center}


To be more illustrative, the set of random variables
${\boldsymbol{p}}_{n}^{(m)}$ can be presented in the form of a
table, where every row appears after every of the $N$ particles
from system $A$ (the left side of the table) and $B$ (the right
side of the table) has collided:
\bigskip

\begin{minipage}[t]{80mm}
\begin{tabular}{|c|c|c|c|c|c|c|}
  \hline
    & $\vphantom{\frac {\frac 12}2}1$ &  \dots & $N$  & $N+1$ & \dots & $2N$ \\
  \hline
         &\!\! ${\bs k}_0$ &  \ldots & ${\bs k}_0$ & $-{\bs k}_0$  & \ldots & $-{\bs k}_0$\\
  1      & ${\bs p}_1^{(1)}$  & \ldots & ${\bs p}_N^{(1)}$ & ${\bs p}_{N+1}^{(1)}$ & \ldots & ${\bs p}_{2N}^{(1)}$\\
  2      & ${\bs p}_1^{(2)}$  & \ldots & ${\bs p}_N^{(2)}$      & ${\bs p}_{N+1}^{(2)}$  & \ldots & ${\bs p}_{2N}^{(2)}$ \\
   \vdots &  \vdots &  \vdots  & \vdots &  \vdots &  \vdots &  \vdots  \\
  M      & ${\bs p}_1^{(M)}$  & \ldots & ${\bs p}_N^{(M)}$ & ${\bs p}_{N+1}^{(M)}$  & \ldots & ${\bs p}_{2N}^{(M)}$\\
  \hline
\end{tabular}\end{minipage}
\bigskip

\noindent
The final momentum ${\boldsymbol{k}}_{1}$ of the first
particle can be determined by summing up quantities in the first
column of the table, the final momentum ${\boldsymbol{k}}_{2}$ of
the second particle by summing up quantities in the second column,
and so on.

The main goal of this work is to determine $f_{2N}$, the density
distribution function in the momentum space, which describes
$2N$ nucleons after $M$ collisions per particle.
We write the  total energy and momentum of the nucleons after freeze-out:
\begin{equation}
E_{\mathrm{tot}}=\sum_{n=1}^{2N}\epsilon _{n}\quad {\mbox{and}}\quad
{\bs P}_{\mathrm{tot}}=\suml_{n=1}^{2N}{\bs k}_{n} \, ,
\label{1}
\end{equation}
where
$\epsilon _n=\omega (\boldsymbol{k}_n)=\sqrt{m^2
+{\boldsymbol{k}}_n^2}$.
All consideration is carried out in the c.m.s. of two identical
colliding nuclei, hence the initial total momentum of the system
$A+B$ is equal to zero.
As we discuss further, the total momentum of
the system of $2N$ nucleons, $\boldsymbol{P}_{\mathrm{tot}}$, can
slightly differ from zero after freeze-out because of creation of
secondary particles.
Let us write down the density distribution function in the form
\begin{equation}
f_{2N}=C\widetilde{f}_{2N}\,,
\label{def-1}
\end{equation}
where $C$ is the normalization constant.
The unnormalized distribution function $\tilde{f}_{2N}$ can be defined
in a two-fold way:
first we follow all collisions of particular nucleon by integration
with respect to all nucleon random momentum transfer,
second we fix the total energy and total momentum of $2N$-nucleon
system after freeze-out in a micro-canonical-like way.
Then, it reads
\[
\widetilde{f}_{2N}(E_{\rm tot}, \bs P_{\rm tot};{\bs k}_1,\ldots
,{\bs k}_{2N} ) =
\]
\[
= \! \intl  \frac{dP_1}{V}\,    \ldots \frac{dP_{2N}}{V}
  \prod_{n=1}^N \left[ \delta^3 \! \left( {\bs k}_n-
{\bs k}_0 - \sum_{m=1}^M {\bs p}_n^{(m)}\right) \right] \, \times
\]
\[
\times \prod_{n=N+1}^{2N} \left[ \delta^3 \left( {\bs k}_n + {\bs
k}_0 - \sum_{m=1}^M {\bs p}_n^{(m)}\right) \right]
\]
\begin{equation}
 \times \, \delta \! \left( E_{\rm tot} - \sum_{n=1}^{2N} \epsilon_n
\right) \, V_p \ \delta^3 \! \left( {\bs P}_{\rm tot} - \sum_{n=1}^{2N}
{\bs k}_n \right) \,
\, ,
\label{A}
\end{equation}
where $V$ is the volume of the system in the coordinate space and
the element of volume in the momentum transfer space
accessible for the $n$-th particle in the series of $M$ collisions
reads
\begin{equation}
dP_n \equiv
J\left({\bf p}_n^{(1)}\right)\, d^3p_n^{(1)}
\, \cdot \, \ldots \, \cdot \,
J\left({\bf p}_n^{(M)}\right)\, d^3p_n^{(M)} \, ,
\label{0}
\end{equation}
where the distribution of the momentum transfer is characterized by
the presence of the form-factor $J({\bf p})$.
For the sake of simplicity we assume that this distribution is
homogeneous one and the available values of the transferred momentum
are restricted from above.
So, as a first approximation we adopt that the form-factor has the
following form:
\begin{equation}
J({\bf p})
=
\frac 1{V_p} \, \theta(p_{\rm max}-|p_x|)\, \theta(p_{\rm max}-|p_y|)
\, \theta(p_{\rm max}-|p_z|)
\, ,
\label{ff}
\end{equation}
where $V_p=8 p_{\rm max}^3$ and $\int d^3p \, J({\bf p})=1$.

If the value of the maximally allowed transferred momentum scales as
$p_{\rm max} \sim 1$~GeV/c then, our assumption actually means that
nucleon-nucleon (hadron-hadron) reactions cannot go in the spatial
volume which size is less than $0.2$~fm or the nucleons cannot be
closer than the distance $0.2$~fm (compare this with the
models where nucleon-nucleon potential has a hard core).

Because we fix the total momentum of the nucleon system after
freeze-out by the presence of the
$\delta$-function, $\delta ^{3}\left( \boldsymbol{P}_{\mathrm{tot}}
-\sum_{n=1}^{2N}{\boldsymbol{k}}_{n}\right) $, we introduce the
multiplier $V_{p}$ just to provide the dimension of the density
distribution in the same way as of common use in statistical
mechanics.

Note, because of creation of secondary particles the energy of the
nucleon system after freeze-out, $E_{\rm tot}$, is in the following
correspondence with the initial energy, $E_0$, which is fixed as the
energy of net nucleons before collision in the c.m.s. of colliding
nuclei
\begin{equation}
E_0=E_{\rm tot} + E_{\rm sp}
\, ,
\label{ee}
\end{equation}
where $E_{\rm sp}$ is the energy of the secondary particles.
And in a similar way the total momentum of the nucleon system after
freeze-out, $\bs P_{\rm tot}$, is in the following
correspondence with the initial momentum, $\bs P_0$, of the nucleons
\begin{equation}
\bs P_0=\bs P_{\rm tot} + \bs P_{\rm sp}
\, ,
\label{pp}
\end{equation}
where $\bs P_{\rm sp}$ is the momentum of the secondary particles.
If we investigate the collisions of identical nuclei then,
$\bs P_0=0$.
However, the value of the momentum of the nucleon system after
freeze-out, $\bs P_{\rm tot}$, can be finite and is fixed by
equation $\bs P_{\rm tot} + \bs P_{\rm sp}=0$.

We normalize the density distribution function in such a way that
simultaneously determines the density of states in the system:
\begin{equation}
\Omega_{2N}=\int d{\tilde k}_1
\ldots  d{\tilde k}_{2N}\, \widetilde{f}_{2N}(E_{\rm tot},
\bs P_{\rm tot};{\bs k}_1,\ldots ,{\bs k}_{2N} ) \, ,
\label{2c}
\end{equation}
where the element of single-particle phase space looks like (in units
of $\hbar $): $d{\tilde{k}}_{n}=V\frac{d^{3}k_{n}}{(2\pi )^{3}}\,.$
Hence, the normalization constant $C$ is equal to
\begin{equation}
C=\frac{1}{\Omega _{2N}(E_{\mathrm{tot}},{\bs P}_{\mathrm{tot}})}\,.
\end{equation}

Making allowance for notation (\ref{0}), the unnormalized
density distribution (\ref{A}) can be written down in the form
\[
 \widetilde{f}_{2N}(E_{\rm tot}, \bs P_{\rm
tot};{\bs k}_1,\ldots  ,{\bs k}_{2N} ) =
\]
\[
= \delta \! \left( \! E_{\rm tot}{-} \suml_{n=1}^{2N} \epsilon_n
\right) V_p\, \delta^3 \! \left( \! {\bs P}_{\rm tot} {-}
\suml_{n=1}^{2N} {\bs k}_n \right)
\]
\[
 \times \! \prod_{n=1}^N
\Biggl[ \int \frac{d^3a_n} {V(2\pi)^3}\, e^{-i{\bs
a}_n\cdot ({\bs k}_n - {\bs k}_0) }
\! \prod_{m=1}^M \! \int_{V_p}
\frac{d^3p^{(\!m\!)}_n}{V_p} \, e^{ i{\bs a}_n\cdot {\bs p}_n^{(m)} }
    \Biggr]
\]
\begin{equation}
\times \! \! \! \prod_{n=N+1}^{2N} \Biggr[ \! \! \int \! \frac{d^3b_n}{V(2\pi)^3}\,
e^{-i{\bs b}_n\cdot ({\bs k}_n + {\bs k}_0) }
\! \! \prod_{m=1}^M  \! \int_{V_p} \! \! \frac{d^3p^{(\!m\!)}_n}{V_p} \,
e^{ i{\bs b}_n\cdot {\bs p}_n^{(m)} } \Biggr] .
\label{A3}
\end{equation}
Here, we presented $\delta $-functions, which correspond to the conservation
of momentum in a series of two-particle collisions, in terms of the Fourier
integrals over the variables $\boldsymbol{a}_{n}$ and $\boldsymbol{b}_{n}$.

Let us define an auxiliary function
\begin{equation}
g({\bs a})\equiv \intl_{V_{p}}\frac{d^{3}p}{V_{p}}\,e^{i{\bs a}\cdot \,{\bs p%
}}=\prod_{i=1}^{3}\frac{\sin {\ \left( a_{i}p_{\mathrm{max}}\right) }}{%
a_{i}p_{\mathrm{max}}}\,,  \label{a-10}
\end{equation}
where the finiteness of the allowed momentum space (determined by (\ref{ff}))
has been taken into account.
Making use of this function, the unnormalized distribution density,
$\tilde{f}_{2N}$, can be rewritten in the unified form
\[
 \widetilde{f}_{2N}(E_{\rm tot}, \bs P_{\rm tot};{\bs k}_1,\ldots
,{\bs k}_{2N} ) =
\]
\[
 = \delta \left( E_{\rm tot}-
\suml_{n=1}^{2N} \epsilon_n \right) V_p \, \delta^3 \left( {\bs
P}_{\rm tot} - \suml_{n=1}^{2N} {\bs k}_n \right)
\]
\[
\times \, \prod_{n=1}^N \left[ \frac{1}{V} \int
\frac{d^3a_n}{(2\pi)^3}\, e^{-i{\bs a}_n\cdot ({\bs k}_n - {\bs
k}_0) + M \ln{ g( {\bs a}_n )} } \right]
\]
\begin{equation}
\times \prod_{n=N+1}^{2N} \left[ \frac{1}{V} \int
\frac{d^3b_n}{(2\pi)^3}\,  e^{-i{\bs b}_n\cdot ({\bs k}_n + {\bs
k}_0) + M \ln{ g( {\bs b}_n )} } \right] \, .
\label{A4}
\end{equation}

To obtain the partition function, we should make the Laplace transformation
of the density-of-states function (\ref{2c}) with respect to the variable
$E_{\mathrm{tot}}$ (it is evident that the function $\Omega _{2N}$ depends on
the external parameters
$E_{\mathrm{tot}}$ and ${\boldsymbol{P}}_{\mathrm{tot}}$):
\[
 Z_{2N}(\beta,{\bs P}_{\rm tot}) =  \intl_{E_{\rm min}}^\infty
dE_{\rm tot} \, e^{ -\beta E_{\rm tot} } \, \Omega_{2N}(E_{\rm
tot},{\bs P}_{\rm tot})=
\]
\[
 = \intl_{E_{\rm min}}^\infty
dE_{\rm tot} \, e^{ -\beta E_{\rm tot} }
\]
\begin{equation}
\times \, \int  d{\tilde k}_1  \ldots d{\tilde k}_{2N} \,
\widetilde{f}_{2N}(E_{\rm tot},\bs P_{\rm tot};{\bs k}_1, \ldots
,{\bs k}_{2N} ) .
\label{a-12}
\end{equation}
This procedure, after the order of integration has been changed,
brings about the Laplace transformation of the unnormalized
density distribution function $\tilde{f}_{2N}$:
\[
 \widetilde {\mathbb{F}}
_{2N} (\beta , \bs P_{\rm tot}; \bs k_1, \ldots, \bs k_{2N}) =
\]
\begin{equation}
 =
\intl_{E_{\rm min}}^\infty dE_{\rm tot} \, e^{ -\beta E_{\rm tot}
} \widetilde{f}_{2N}(E_{\rm tot}, \bs P_{\rm tot}; \bs k_1,
\ldots, \bs k_{2N}) \, ,
 \label{a-11}
\end{equation}
Taking Eq.~(\ref{A4}) into account, we obtain
\[
\widetilde {\mathbb{F}}_{2N}(\beta, \bs P_{\rm tot};
{\bs k}_1,\ldots ,{\bs k}_{2N} )
= V_p \,\delta^3 \left( {\bs P}_{\rm tot} - \suml_{n=1}^{2N} {\bs
k}_n \right)
\]
\[
{\times}  \prod_{n=1}^N  \left[ e^{ -\beta \epsilon_n }
\frac{1}{V} \int  \! \frac{d^3a_n}{(2\pi)^3}\,e^{-i{\bs a}_n\cdot
({\bs k}_n - {\bs k}_0) + M \ln{ g( {\bs a}_n )} } \right]
\]
\begin{equation}
\times \! \prod_{n=N+1}^{2N} \! \! \left[ e^{ -\beta \epsilon_n }
\frac{1}{V} \int \! \! \frac{d^3b_n}{(2\pi)^3}\, e^{-i{\bs b}_n\cdot
({\bs k}_n + {\bs k}_0) + M \ln{ g( {\bs b}_n )} } \right] \! .
\label{A5}
\end{equation}
While doing the Laplace transformation, we did not consider the
possible dependence of $p_{\mathrm{max}}$ on $E_{\mathrm{tot}}$.
Therefore, in what follows, the value of $p_{\mathrm{max}}$ will be
taken constant, thus playing the role of a parameter.

So, the partition function (\ref{a-12}), which is analogue of the
partition function in the canonical ensemble (in what follows, for
simplicity we name it as the canonical ensemble), can be written now
in the following way
\begin{equation}
Z_{2N}(\beta,{\bs P}_{\rm tot}) \! = \!
\int \! d{\tilde k}_1 \ldots d{\tilde k}_{2N}
\widetilde {\mathbb{F}}_{2N}(\beta, \bs P_{\rm tot};
{\bs k}_1,\ldots ,{\bs k}_{2N} )  .
\label{b-12}
\end{equation}
Following this line one can determine the distribution function in
the canonical ensemble
\begin{equation*}
{\mathbb{F}}_{2N}\left(\beta, \bs P_{\rm tot};{\bs k}_1,\ldots ,
{\bs k}_{2N}\right) \!
=
\frac {\widetilde {\mathbb{F}}_{2N}(\beta, \bs P_{\rm tot};
{\bs k}_1,\ldots ,{\bs k}_{2N} ) }{Z_{2N}(\beta,{\bs P}_{\rm tot})}
\, .
\label{c-12}
\end{equation*}
As a matter of fact, the mean values of the physical quantities which
are evaluated with the help of the distribution function,
$f_{2N}=\widetilde{f}_{2N}/\Omega _{2N}$ from (\ref{def-1}),
are obtained when the total energy of nucleons after freeze-out,
$E_{\rm tot}$, is fixed.
Meanwhile, the mean values evaluated with making use of the
distribution function,
${\mathbb{F}}_{2N}(\beta,\bs P_{\rm tot};{\bs k}_1,\ldots,{\bs k}_{2N})$,
will be at fixed parameter $\beta$.

\section{Partition Function for a Large Number of Collisions}
To calculate integrals over the variables $\boldsymbol{a}_{n}$ and
$\boldsymbol{b}_{n}$ on the r.h.s. of Eq.~(\ref{A5}) in the case
where the number of collisions for every particle is large, $M\gg 1$, the
saddle-point method can be applied.
It implies the use of the following approximation:
\begin{equation}
e^{ M \ln{ g( {\bs a} )}}
= e^{M \suml_{i=1}^3 \ln{ \left(
\frac{\sin{ \left( a_i p_{\rm max} \right) } } { a_i p_{\rm max}
}\right) } } \approx e^{ -\frac{M}{6} {\bs a}^2 p_{\rm max}^2 } \, .
\label{10}
\end{equation}
This expression is to be substituted to (\ref{A5}).
The analogous operation is fulfilled in relation to the integration
variables ${\boldsymbol{b}}_{n}$.
Using the Poisson integral we make integration in (\ref{A5})
with respect to the variables ${\boldsymbol{a}}_{n}$ and
${\boldsymbol{b}}_{n}$.
Then, the partition function (\ref{b-12}) for the canonical ensemble,
can be written in the following way
\begin{equation}
Z_{2N}(\beta,{\bs P}_{\rm tot}) = \frac{V_p}{(2\pi)^3} \int d^3x \,
e^{-i{\bs P}_{\rm tot}\cdot {\bs x}}
\label{12}
\end{equation}
\[
\times \prod_{n=1}^N \left[ \left(  \frac{\alpha}{\pi} \right)^{3/2}
\!\!\int \frac{d^3k_n}{(2\pi)^3} \, e^{ -\beta \omega({\bs k}_n) + i
{\bs k}_n \cdot {\bs x} - \alpha({\bs k}_n - {\bs k}_0)^2} \right]
\, \times
\]
\[
 \times \!\!\prod_{n=N+1}^{2N}  \!\!
 \left[ \left( \frac{\alpha }{\pi} \right)^{3/2} \! \!
 \int \frac{d^3k_n}{(2\pi)^3} \,e^{ -\beta \omega({\bs k}_n) +
 i {\bs k}_n \cdot {\bs x} - \alpha({\bs k}_n +
{\bs k}_0)^2  }\right] \, ,
\]
where the $\delta $-function, which reflects the conservation law of
total momentum of the nucleon system after freeze-out, is presented in
the form of Fourier integral over the variable
${\boldsymbol{x}}$, and the parameter $\alpha$ is defined as:
\begin{equation}
\dis\alpha \equiv \frac{3}{2Mp_{\mathrm{max}}^{2}}
\,.
\label{12prime}
\end{equation}
Note, if instead of the Cartesian coordinate system one makes
evaluations in the spherical coordinate system in
the space of momentum transfer, then instead of the form-factor
(\ref{ff}) and the quantity $\alpha$ defined in (\ref{12prime}) one
should use expressions:
$J({\bf p})= \theta(p_{\rm max}-|\bs p|)/V_p$, where
$V_p=4 \pi p_{\rm max}^3/3$ and $\alpha =5/(2Mp_{\mathrm{max}}^2)$,
respectively.

We define auxiliary single-particle functions
\[
z_{a(b)}(\beta,{\bs x}) \equiv
\left( \frac{\alpha}{\pi}
\right)^{3/2} \int \frac{d^3k}{(2\pi)^3}
\]
\begin{equation}
\times \exp{ \left[ -\beta \omega({\bs k}) + i {\bs k} \cdot {\bs x}
- \alpha({\bs k} \mp {\bs k}_0)^2 \right] } \label{14}
\end{equation}
for the systems (nuclei) A and B, respectively. With their help,
the partition function $Z_{2N}(\beta ,{\boldsymbol{P}}_{%
\mathrm{tot}})$ from (\ref{12}) can be rewritten as follows:
\begin{equation}
Z_{2N}
 {=} \frac{V_p}{(2\pi)^3} \int \!\! d^3x \,  e^{ {-}i
{\bs P}_{\rm tot} \cdot {\bs x} {+} N \ln{z_a(\beta,{\bs x})} {+} N
\ln{z_b(\beta,{\bs x})} } . \label{15}
\end{equation}
This expression will be basic for further research in this work.

\subsection*{3.1. Collision of two identical nuclei}

In the case of collision between two identical nuclei,  all
nucleons in the one nucleus have the initial momentum
${\boldsymbol{k}}_0$
and all nucleons in the other nucleus have the initial momentum
$-{\boldsymbol{k}}_0$.
The calculation of partition function
(\ref{15}) will be carried out making use of the saddle-point method
at $N\gg 1$.
We expand the functions $\ln {z_a(\beta
,}\boldsymbol{x}{)}$ and $\ln {z_b(\beta ,\boldsymbol{x})}$ (see
(\ref{14})) into series with respect to the variable
${{\boldsymbol{x}}}$ about the point ${{\boldsymbol{x}}}=0$ up to
the second order inclusive:
\vspace{-1mm}
\[
 \ln{z_a(\beta,{\bs x})} {\approx}
\ln{z_a(\beta,0)} {+} \!\sum_{i=1}^3 \! \left[\!\!
\frac{1}{z_a(\beta,{\bs x})} \frac{\partial z_a(\beta,{\bs
x})}{\partial x_i} \right]_{ {\bs x}=0 } \! x_i
\]\vspace{-1mm}
\[
+ \frac{1}{2} \suml_{i,j=1}^3 \Biggl[ \frac{1}{z_a(\beta,{\bs x})}
\frac{\partial^2 z_a(\beta,{\bs x})}{\partial x_j
\partial x_i}
\]
\begin{equation}
 - \frac{1}{z_a^2(\beta,{\bs x})} \frac{\partial
z_a(\beta,{\bs x})}{\partial x_j } \frac{\partial z_a(\beta,{\bs
x})}{\partial x_i } \Biggr]_{ {\bs x}=0 } x_j x_i \, . \label{101}
\end{equation}
We define two functions (see (\ref{14}))
\begin{equation}
z_a(\beta) \equiv z_a(\beta,0) \, , \ \ \ \ z_b(\beta) \equiv
z_b(\beta,0) \, . \label{103}
\end{equation}
Then, regarding $z_{a}(\beta )$ and $z_{b}(\beta )$ as the
single-particle partition functions, the terms in brackets on the
r.h.s. of (\ref{101}) can be taken as the statistical averages of
$k_{i}$, $k_{i}k_{j}$, and so on. Indeed, we designate such
quantities by angle parentheses $\left\langle \ldots \right\rangle
_{a}$ and $\left\langle \ldots \right\rangle _{b}$, where the
subscripts $a$ and $b$ mean that averaging is carried out making use
of the single-particle partition function $z_{a}(\beta )$ or
$z_{b}(\beta )$, respectively:
\begin{equation}
\langle \ldots \rangle_{a(b)} \equiv \frac { \left( \alpha/\pi
\right)^{3/2}}{z_{a(b)}(\beta)} \int \!\!\!\frac{d^3k}{(2\pi)^3} (
\ldots )\, e^{  {-}\beta \omega({\bs k}) {-} \alpha({\bs k} \mp {\bs
k}_0)^2  } . \label{104}
\end{equation}
Thus, we can rewrite expression (\ref{101}) in terms of the averaged
momentum components:
\vspace{-2mm}
\[
\ln{z_a(\beta,{\bs x})} \approx \ln{z_a(\beta)}
 + i \suml_{i=1}^3 \langle k_i \rangle_a x_i \, -
\]
\vspace{-2mm}
\begin{equation}
   -\frac{1}{2} \suml_{i,j=1}^3 \Big( \langle k_i k_j \rangle_a - \langle
k_i \rangle_a \langle k_j \rangle_a  \Big) x_i x_j \, .
\label{102}
\end{equation}
It is evident that the analogous expression can be written down for
particles from system $B$ as well.

Now, we introduce the correlation function of momenta,
\begin{equation} C_{ij}^{(r)} \equiv \,  \langle k_i k_j \rangle_r
- \langle k_i \rangle_r \langle k_j \rangle_r  \, , \label{105}
\end{equation}
where $r=a$ or $b$, and write expression (\ref{102}) for system $A$ and
the analogous one for system $B$:
\vspace{-2mm}
\[\ln{z_r(\beta,{\bs x})} \approx \ln{z_r(\beta)}
+ i \langle {\bs k} \rangle_r \cdot {\bs x} \, - \frac{1}{2}
\suml_{i,j=1}^3 C_{ij}^{(r)} \, x_i x_j \, . \]
\vspace{-1mm}
It allows us to rewrite partition function (\ref{15}) as follows:
\[
Z_{2N}(\beta) \approx z_a^N(\beta) \, z_b^N(\beta) \,
\frac{V_p}{(2\pi)^3}  \int d^3 x
\]\vspace{-5mm}
\begin{equation}
\times \exp{ \left[ {-}i {\bs P}_{\rm tot} \cdot {\bs x} {+} i \, 2 N
\langle {\bs k} \rangle \cdot {\bs x} {-} N \suml_{i,j=1}^3 C_{ij}
x_i x_j \right] }  ,
\label{107}
\end{equation}
where
\begin{equation}
\label{107prime}
C_{ij}= \frac 12 \left(
C_{ij}^{(a)}+C_{ij}^{(b)}\right)\,, \quad \langle {\bs k} \rangle
= \frac 12 \Big( \langle {\bs k} \rangle_a + \langle {\bs k}
\rangle_b \, \Big)\,.
\end{equation}
Integration of the Poisson integral on the r.h.s. of
Eq.~(\ref{107}) brings about a simpler result
\[
Z_{2N}(\beta) \, \approx \, z_a^N(\beta) \, z_b^N(\beta)
\left( \frac{1}{4\pi N} \right)^{3/2} \!
\frac {V_p}{\left(\det{\widehat C}\right)^{1/2}} \ \times
\]
\begin{equation}
 \times \exp{\left[ -N \left( {\bs p}_{\rm tot}-  \langle
{\bs k} \rangle \right) \cdot {\widehat C} ^{-1}\cdot \left( {\bs
p}_{\rm tot}- \langle {\bs k} \rangle \right)\right] }
 \, ,
 \label{109}
 \end{equation}
where
${\boldsymbol{p}}_{\mathrm{tot}}=\boldsymbol{P}_{\mathrm{tot}}/2N$.

From expression (\ref{109}), one can determine the
``two-source'' single-particle partition function,
which simultaneously concerns both systems $A$ and $B$:
\[
z(\beta) = z_a(\beta)\, z_b(\beta) \left( \frac{1}{4\pi N}
\right)^{3/2N} \! \! \left( \frac {V_p}{\left( \det{\widehat C}
\right)^{1/2}} \right)^{1/N} \!\!\!
\]
\begin{equation}
\times \exp \left[ - \left( {\bs p}_{\rm tot}-  \langle {\bs k}
\rangle \right) \cdot {\widehat C} ^{-1}\cdot \left( {\bs p}_{\rm
tot}- \langle {\bs k} \rangle \right)  \right] \, , \label{110}
\end{equation}
so that we can write down that
\begin{equation}
Z_{2N}(\beta )=z^{N}(\beta )\,.  \label{111}
\end{equation}

We would like to emphasize the validity of the following limit value:
\begin{equation*}
\lim_{N\rightarrow \infty } \left(\frac 1{4\pi N}\right)^{3/2N}=1\,.
\end{equation*}
For instance, at $N=10,$ 100, 200, and 1000, the corresponding estimations are
$(1 /(4\pi N))^{3/2N}=0.48$, 0.90, 0.94, and 0.99.
As a result, in the case ${%
\boldsymbol{P}}_{\mathrm{tot}}=0$ and provided that the particle number $N$
is rather large, it follows from Eq.~(\ref{110}) that
\begin{equation*}
\lim_{N\rightarrow \infty }z(\beta )
=z_{a}(\beta )\,z%
_{b}(\beta )\,e^{-\langle {\bs k}\rangle \cdot {\widehat{C}}^{-1}\cdot
\langle {\bs k}\rangle }\,.
\end{equation*}
%

\subsection*{3.2. Calculation of correlation matrix
                  \boldmath{$C_{ij}$}}

Consider now the calculation of the elements of correlation matrix for the
collision of two identical nuclei in the center-of-mass frame, i.e. we adopt
the initial momentum $\boldsymbol{k}_{0}=(0,0,k_{0z})$.

We would like to recall the definition of the correlation matrix,
\[
 C_{ij} \equiv \frac 12 \left[
C_{ij}^{(a)}+C_{ij}^{(b)} \right] =
\]
\begin{equation}
 = \frac
12 \left[ \vphantom{C_{ij}^{(a)}} \langle k_i k_j\rangle_a +
\langle k_i k_j \rangle_b - \langle k_i \rangle_a \langle k_j
\rangle_a - \langle k_i \rangle_b \langle k_j \rangle_b \right]
\,.
 \label{114}
\end{equation}
We note, the average of transverse components of the
momentum is equal to zero,
$\left\langle k_{x}\right\rangle _{a(b)}=\left\langle
k_{y}\right\rangle _{a(b)}=0$.
From definition (\ref{104}) it is evident
that the effective average momentum (\ref{107prime}) is
\[
\langle {\bs k} \rangle \equiv \frac 12 \left( \vphantom{ \frac
12} \langle {\bs k} \rangle_a +\langle {\bs k} \rangle_b \right) =
\left(0,0, \frac 12\left( \langle k_z \rangle_a +\langle k_z
\rangle_b \right) \right)\,,
\]
while
\begin{equation}
\langle  k_x k_y \rangle_{a(b)}
=\langle  k_x k_z \rangle_{a(b)} =\langle  k_y k_z \rangle_{a(b)}
=0  \,,
\label{114b}
\end{equation}
i.e. all the non-diagonal elements of the correlation matrix are equal to
zero.
As a result, the matrices $\hat{C}^{(a)}$ and $\hat{C}^{(b)}$ are
diagonal; hence, the matrix inverse to $\hat{C}$ is diagonal as well,
namely,
\[
{\widehat C}^{-1}{=} \left( \begin{array}{ccc} \dis \frac{2}{C_{11}^{(a)}
{+} C_{11}^{(b)}} & 0 & 0 \\ 0 &  \dis  \frac{2}{C_{22}^{(a)} {+}
C_{22}^{(b)}} & 0 \\ 0 & 0 &  \dis \frac{2}{C_{33}^{(a)} {+}
C_{33}^{(b)}} \\ \end{array}  \right).
\]

Consider the two-source single-particle partition function (\ref{110})
in the case where the total momentum of the nucleon sub-system after
freeze-out equals to zero,
${\boldsymbol{p}}_{\mathrm{tot}}=\boldsymbol{P}_{\mathrm{tot}}/2N=0$.
Bearing the structure of the
correlation matrix in mind and taking the value of the average moment $%
\left\langle {\boldsymbol{k}}\right\rangle $ into account, partition
function (\ref{110}) can be rewritten in the form
\[
 z(\beta) = z_a(\beta)\, z_b(\beta) \left(
\frac 1{2\pi N} \right)^{3/2N}
\]
\begin{equation}
\times \left[ \frac{V_p}{ \prod_{i=1}^3
\left(C_{ii}^{(a)}{+}C_{ii}^{(b)}\right)^{1/2}} \right]^{1/N} \,
e^{ -\, C_{zz}^{-1} \langle k_z \rangle^2 } \,   .
\label{115}
\end{equation}
On the other hand, the $z$-component of the average momentum is equal to
\[
\langle k_z \rangle
= \frac12 \Bigl( \vphantom{ \frac 12} \langle k_z \rangle_a
 +\langle k_z \rangle_b \Bigr)  =
\]
\[
= \frac{\left(\alpha/\pi \right)^{3/2} }{2z_a(\beta)} \int
\frac{d^3k}{(2\pi)^3}  k_z \exp \left[ -\beta \omega({\bs k}) -
\alpha({\bs k}- {\bs k}_0)^2 \right]
\]
\begin{equation}
 {+} \frac{\left(\alpha/\pi \right)^{3/2}}
{2z_b(\beta)} \int \frac{d^3k}{(2\pi)^3} \, k_z\, \exp{\left[
{-}\beta \omega({\bs k}) {-} \alpha({\bs k}{+} {\bs k}_0)^2 \right]
} . \label{116}
\end{equation}
We notice
that the change of the integration variable $k_{z}\rightarrow
-k_{z}$ in integral (\ref{14}) at $\boldsymbol{x}=0$ makes it
evident that $z_{a}(\beta )=z_{b}(\beta )$.
Using this equality, we obtain, from Eq.~(\ref{116}), that
\begin{equation*} \langle k_z \rangle \sim
\intl_{-\infty}^{\infty} \frac{dk_z}{2\pi} \, k_z\, e^{-\beta
\omega({\bs k})} \left[  e^{ - \alpha(k_z- k_{0z})^2 } {+}  e^{ -
\alpha(k_z + k_{0z})^2  } \right]\, .
\label{118}
\end{equation*}
Here, the integrand
is no more than a product of an even and an odd functions.
Owing to the symmetry of integration limits, the result of
integration vanishes.
At last, we have $\left\langle k_{z}\right\rangle =%
\frac{1}{2}\left( \left\langle k_{z}\right\rangle _{a}+\left\langle
k_{z}\right\rangle _{b}\right) =0$, i.e.
$\left\langle k_{z}\right\rangle _{a}=-\left\langle k_{z}\right\rangle _{b}$.
This result was expectedly obtained for identical nuclei when
$\boldsymbol{P}_{\mathrm{tot}}$ equals zero.

Taking into account that $z_{a}(\beta )=z_{b}(\beta )$ and the
matrices $\hat{C}^{(a)}$ and $\hat{C}^{(b)}$ are diagonal, we draw a
conclusion that those matrices coincide, i.e.
\begin{equation}
{\widehat{C}}^{(a)}={\widehat{C}}^{(b)}={\widehat{C}}\,.
\label{119a}
\end{equation}%
As a consequence, we can simplify Eq.~(\ref{115}) and write down the
ultimate expression for the single-particle partition function:
\[
 \lim_{N \to \infty}z(\beta)  =  z_a(\beta)\,
z_b(\beta) =
\]
\[
 =  \dis \left( \frac{\alpha}{\pi }\right)^3
\int \frac{d^3k_a}{(2\pi)^3} \exp{ \left[ -\beta
\omega({\bs k}_a) - \alpha({\bs k}_a -{\bs k}_0)^2  \right] }
\]
\begin{equation}
 \times  \int\frac{d^3k_b}{(2\pi)^3} \exp{
\left[ -\beta \omega({\bs k}_b)- \alpha({\bs k}_b + {\bs k}_0)^2
\right] } \, ,
\label{120}
\end{equation}
where the approximation
\begin{equation*}
\left( \frac 1{4\pi N}\right)^{3/2N}\left( \frac{V_p}{\left( \det
{\widehat{C}}\right) ^{1/2}}\right)^{1/N}\approx 1,
\end{equation*}%
fair at large enough $N$, was taken into account.

Hence, on the basis of the proposed approach, the following
nonequilibrium ``two-source'' distribution function has been obtained:
\[
 {f}({\bs k}_a, {\bs k}_b)
 {=} \frac{\left(\alpha/\pi \right)^3 }{z_a(\beta)z_b(\beta)}
 \exp{ \left[ -\beta \omega({\bs k}_a)
   {-} \alpha({\bs k}_a -{\bs k}_0)^2 \right] }
\]
\begin{equation}
\times \exp{ \left[ -\beta \omega({\bs k}_b) - \alpha({\bs k}_b +
{\bs k}_0)^2  \right] } \, , \label{120prime} \end{equation}
where in accordance with (\ref{103})
\begin{equation}
z_{a(b)}(\beta) = \left(\frac{\alpha}{\pi}  \right)^{3/2}
\int \frac{d^3k}{(2\pi)^3} \,
e^{-\beta \omega({\bs k})- \alpha ({\bs k} \mp {\bs k}_0)^2 }
\, .
\label{122-9a}
\end{equation}
One can see that two-source distribution function (\ref{120prime})
demonstrates
features associated with the central limit theorem, which manifest
themselves in the availability of two Gaussians, symmetrically located
in the momentum space (remind, the analysis is carried out in the c.m.s.
of two colliding nuclei).
Actually, the latter is the reason for the name ``two-source''.
The proposed approach allowed
not only the expected general result to be obtained, but also the
expression for the dispersion and the distribution centers to be
deduced, which can be checked up experimentally in nucleus-nucleus
collisions.

In particular, we note that the increase of the number of collisions
$M$ and the accessible volume in the momentum transfer space, which is
determined by the
parameter $p_{\mathrm{max}}$ (see (\ref{12prime})), gives rise to the
smearing of this effect and the transformation of the distribution
into the thermal one.

\subsection*{3.3. Single-particle spectrum}

Consider the spectrum of particles, $d^3 N/d^3p$, after freeze-out.
If every nucleon from the large system, according to the
proposed model, takes part in $M$ collisions, the spectrum
can be obtained by averaging the quantity
(as well as an arbitrary function
$D(\boldsymbol{k}_{a},\boldsymbol{k}_{b})$ of random variables)
\begin{equation}
D(\bs p,{\bs k}_{a},{\bs k}_{b})=N\delta ^{3}({\bs p}-{\bs k}_{a})+N\delta
^{3}({\bs p}-{\bs k}_{b})
\label{122-9aprime}
\end{equation}
over the values of ${\boldsymbol{k}}_{a}$ and ${\boldsymbol{k}}_{b}$ with
making use of two-source single-particle distribution function
(\ref{120prime}), i.e.
\begin{equation}
 \langle D \rangle
=
\int \frac{d^3k_a}{(2\pi)^3} \, \frac{d^3k_b}{(2\pi)^3} \,
D(\bs p,{\bs k}_a,{\bs k}_b) \, {f}({\bs k}_a, {\bs k}_b)
 \, ,
\label{122-7}
\end{equation}
Then, after averaging the random quantity (\ref{122-9aprime}), the
spectrum, $d^3 N/d^3p=\big\langle D \big\rangle$, looks like
\[
\frac{d^3 N}{d^3p}
= N \left(\frac{\alpha}{\pi } \right)^{3/2} e^{-\beta \omega({\bs p})}
\, \times
\]
\begin{equation}
\times \left[ \frac{1}{ z_a(\beta) } \,
  e^{ - \alpha({\bs p}- {\bs k}_0)^2  }
+ \frac{1}{ z_b(\beta) } \,
  e^{ - \alpha({\bs p}+ {\bs k}_0)^2   } \right]
\label{301}
\end{equation}
with $\alpha$ defined in (\ref{12prime}).
The spectrum (\ref{301}) is obtained as a sum of two terms which
represent the two-source structure: the first contribution
$\propto \exp{ [-\beta \omega({\bs p})- \alpha({\bs p}-
{\bs k}_0)^2] } /z_a$ is associated with nucleus ``A'', and the second
$\propto \exp{ [-\beta \omega({\bs p})- \alpha({\bs p}+
{\bs k}_0)^2] } /z_b$ is associated with nucleus ``B''.

If we consider the initial momentum $\bs{k}_{0}=(0,0, k_{0z})$,
the spectrum possesses two maxima near the two values of the
longitudinal momenta: $p_{z}=k_{0z}$ and $p_{z}=-k_{0z}$.
Let the notation $\boldsymbol{p}_{\bot }^{2}$ stands for
$p_{x}^{2}+p_{y}^{2}$; then, the particle spectrum in the momentum
space looks like
\[ \frac{d^3 N}{d^3p}
=  \frac {N \left(\alpha/\pi  \right)^{3/2}} { z_0(\beta)}\, \ e^{
-\beta \omega({\bs p})  - \alpha {\bs p}_\bot^2 }  \, \times
\]
\begin{equation}
\times \left[ e^{ - \alpha(p_z-k_{0z})^2  } + e^{ -
\alpha(p_z+k_{0z})^2 } \right] \, , \label{122-10} \end{equation}
where
\begin{equation}
z_0(\beta) = \left(\frac{\alpha}{\pi } \right)^{3/2}
\int \frac{d^3k}{(2\pi)^3} \, e^{ -\beta \omega({\bs k})
 - \alpha \left[ {\bs k}_\bot^2+(k_z-k_{0z})^2 \right] }
\, .
\label{122-12}
\end{equation}
%

\subsection*{3.4. Effective temperature at $\bs k_0=0$}

It is worth noting that in kinetic approach, where the distribution
function depends on time, $f(t,\bs k)$, the temperature can
be determined just conventionally.
The latter problem is analogous to a definition of instant velocity of
the object, which moves with acceleration.
Actually, in our approach we parametrize the time axis by the number
of particle collisions $M=\langle \nu \rangle t$, where
$\langle \nu \rangle$ is the mean frequency of collisions and $t$ is
the time interval.
In this sense the obtained distribution function (\ref{120prime}) with
respect to variable $M$ can be regarded as a nonequilibrium one
($M \propto$ time) and the parameter $\beta=1/T$ cannot be regarded as
an inverse temperature.
Just in the asymptotic regime when $t\to \infty$ or on our language
$M\to \infty$ the distribution function (\ref{120prime}) transforms
to the Boltzmann one and the parameter $T$ becomes the Boltzmann
temperature.
Note, for the sake of simplicity we do not consider special statistics
so far.

Model, which describes a subsystem of particles with a zero initial
momentum, $\bs k_0=0$, can be applied as a first approximation
to the secondary particles, for instance pions, which are created in
relativistic nucleus-nucleus collision.
One can regard the momentum of the particle just after creation as a
first momentum transfer, which is added to zero initial momentum.
On the other hand, a particular example discussing in the present
paragraph can be regarded as a mathematical limit when
$\bs k_0 \to 0$.

First, consider particles, which are created near threshold.
In this case a nonrelativistic behavior is relevant to single-particle
energy
$\omega ({\boldsymbol{p}})\approx m+{\boldsymbol{p}}^{2}/2m$, where
$m$ is the particle's mass.
The subsystem under consideration consists of $N$ particles:
for condition $\bs k_0=0$ the distribution function (\ref{120prime})
reduces to
\begin{equation}
f({\bs k})
= \frac{\left(\alpha/\pi \right)^3 }{z(\beta)}
\exp{ \left[ -\beta \omega({\bs k})- \alpha {\bs k}^2 \right] }
\, ,
\label{df0}
\end{equation}
where
\begin{equation}
z(\beta) = \left(\frac{\alpha}{\pi}  \right)^{3/2}
\int \frac{d^3k}{(2\pi)^3} \,
\exp{\left[ -\beta \omega({\bs k})- \alpha {\bs k}^2 \right]}
\, .
\label{sppf}
\end{equation}
Evidently, for this initial condition the two-source
distribution (\ref{120prime}) reduces to the ``single-source'' one.

An estimation for the effective temperature can be obtained in
the following intuitive way:
\begin{equation}
\int \frac{d^3k}{(2\pi)^3}
e^{ -\frac{{\bs k}^2}{ 2mT} - \alpha {\bs k}^2 }
= \left(\frac{m T_{\rm eff}}{2\pi} \right)^{3/2}
\, , \label{122-1}
\end{equation}
where
\begin{equation}
\frac{1}{T_{\mathrm{eff}}}=\frac{1}{T}+\frac{1}{T_{\mathrm{coll}}}\,,
\label{122-2}
\end{equation}
and the collision \textquotedblleft temperature\textquotedblright\
is defined as
\begin{equation}
T_{\mathrm{coll}}\equiv \frac{Mp_{\mathrm{max}}^{2}}{3m}\,.  \label{122-3}
\end{equation}%
For the thermal wave length determined as
$\Lambda =\sqrt{2\pi /mT}$ and, respectively,
$\Lambda_{\mathrm{eff}}=\sqrt{2\pi /mT_{\mathrm{eff}}}$, the spectrum can
be written down in the form
\begin{equation}
\dis\frac{d^{3}N}{d^3p}
=
\frac{N\,\Lambda _{\mathrm{eff}}^3}{(2\pi )^3\, } \,
\exp{\left[ -\frac{{\bs p}^{2}}{2mT_{\mathrm{eff}}} \right]}
\,.
\label{122-4}
\end{equation}

We conclude immediately: the restriction of the accessible volume in
the space of transferred momentum and the finiteness of the collision
number turn out to effectively reduce the temperature,
$T_{\mathrm{eff}}\leq T$.
Really, according to Eq.~(\ref{122-2}),
\begin{equation}
T_{\mathrm{eff}}=\frac{T}{1+T/T_{\mathrm{coll}}}\leq T\,.
\label{122-5}
\end{equation}
Thus, the increase of the collision number $M$ is accompanied by the
growth of the effective temperature $T_{\mathrm{eff}}$ up to its limit
value $T$.
It is the reason of why, when the parameter $M$ is large enough, the
quantity $T_{\mathrm{eff}}$ in Eq.~(\ref{122-4}) should be replace by
$T$.
On the other hand, if $p_{\mathrm{max}}$ tends to infinity, spectrum
(\ref{122-4}) acquires the standard form
\begin{equation}
\lim_{p_{\mathrm{max}}\rightarrow \infty }\frac{d^3N}{d^3p}=\frac{%
N\,\Lambda ^3}{(2\pi )^3}\,e^{-\frac{{\bs p}^2}{2mT}}\,.
\label{122-6}
\end{equation}

In the general case ($T\sim m$), the relativistic dispersion
relationship $\omega (\mathbf{p})=\sqrt{m^{2}+\mathbf{p}^{2}}$ has to
be used.
Let us make the inverse Laplace transformation of the many-particle
partition function:
\begin{equation}
\Omega _{N}(E_{\mathrm{tot}})
=
\frac1{2\pi i} \intl_{c{-}i\infty }^{c{+}%
i\infty }\hspace{-1mm}d\beta \,e^{\beta
E_{\mathrm{tot}}}\,Z_{N}(\beta )\,.
\label{201}
\end{equation}
In order to calculate the integral on the r.h.s., we use
the saddle-point method.
For this purpose, the last expression is rewritten in the form
\begin{equation}
\Omega _{N}(E_{\mathrm{tot}})
=
\frac1{2\pi i} \intl_{c-i\infty }^{c+i\infty }\,d\beta \,e^{F(\beta )}\,,
\label{202}
\end{equation}%
where Eq.~(\ref{111}) was used and we introduce notation
$F(\beta )=\beta E_{\mathrm{tot}}+N\log{z(\beta )}$.
Now, the minimum of the function
$F(\beta )$ along the imaginary axis and its maximum along the
real axis of the variable $\beta $ are to be determined.
The condition for the extremum to take place along the real axis
(the variable $\beta $ is real) looks like
\begin{equation}
\frac{1}{N}\,E_{\mathrm{tot}}=-\frac{1}{z(\beta )}\,
\frac{\partial z(\beta )}{\partial \beta }\,.
\label{203}
\end{equation}
The solution of this equation with respect to $\beta $ gives the value
of the parameter $T=1/\beta $, which corresponds to the
average energy value per particle, namely, $E_{\mathrm{tot}}/N$.
Here, we may use the explicit form of the single-particle partition
function (\ref{sppf}) to obtain
\begin{equation}
\frac{1}{N}\,E_{\mathrm{tot}}
=
\frac{(\alpha/\pi  )^{3/2}}{z(\beta )}
\int \frac{d^{3}k}{(2\pi )^{3}} \, \omega ({\bs k}) \,
e^{ -\beta \omega ({\bs k}){-}\alpha {\bs k}^{2}} \, .
\label{204}
\end{equation}
We obtained a transcendental equation with respect to the parameter
$\beta $.
The value od this parameter, which is the root of equation, differs
from the value, which is determined from the similar equation for the
ideal gas
\begin{equation}
\frac{1}{N}\,E_{\mathrm{tot}}=\frac{V}{z_{\mathrm{id}}(\beta )}\int
\frac{d^{3}k}{(2\pi )^{3}}\,\omega ({\bs k})\,
e^{-\beta \omega ({\bs k})}\,,
\label{ideal-gas}
\end{equation}
where
\begin{equation}
z_{\mathrm{id}}(\beta )=V\,\int \frac{d^{3}k}{(2\pi )^{3}}\,
e^{-\beta \omega({\boldsymbol{k}})} \, ,
\end{equation}
is the single-particle partition function for the ideal gas.

Similarly to what was done in the non-relativistic case, one can
introduce the idea of effective temperature
$T_{\mathrm{eff}}=1/\beta_{\rm eff}$ in the following way:
$$-\frac 1{z(\beta )} \frac{\partial z(\beta )}{\partial \beta}=
-\frac 1{z_{\rm id}(\beta_{\rm eff} )}
\frac{\partial z_{\rm id}(\beta_{\rm eff} )}{\partial \beta_{\rm eff} }
\, ,$$
where in the non-relativistic case the equality is valid when
$\beta_{\rm eff}=\beta + 1/T_{\rm coll}$.
I.e., by definition the effective temperature, is the temperature
of ideal gas,  at which
%
%
  \begin{center}
  \begin{minipage}{0.9\columnwidth}
\includegraphics[width=0.85\textwidth]{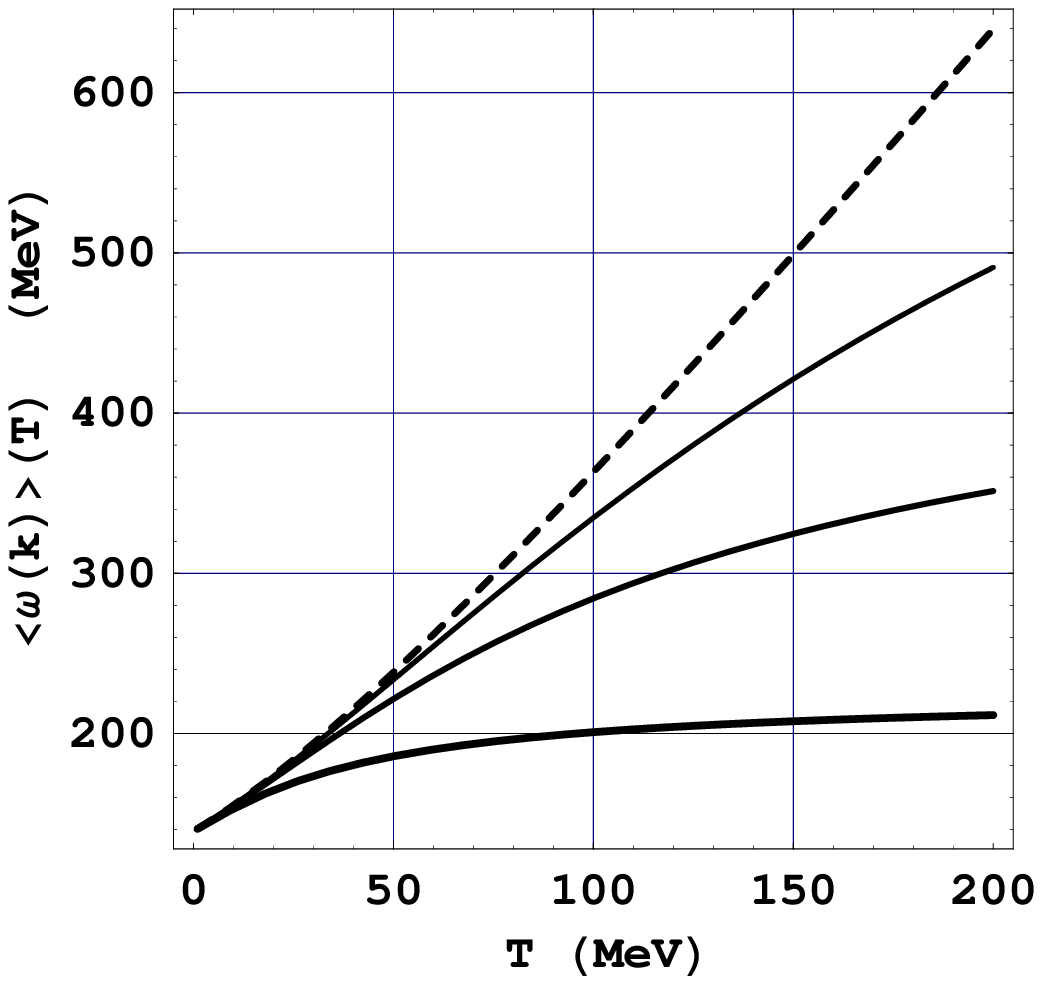}

\vspace{-2mm}
\includegraphics[width=0.85\textwidth]{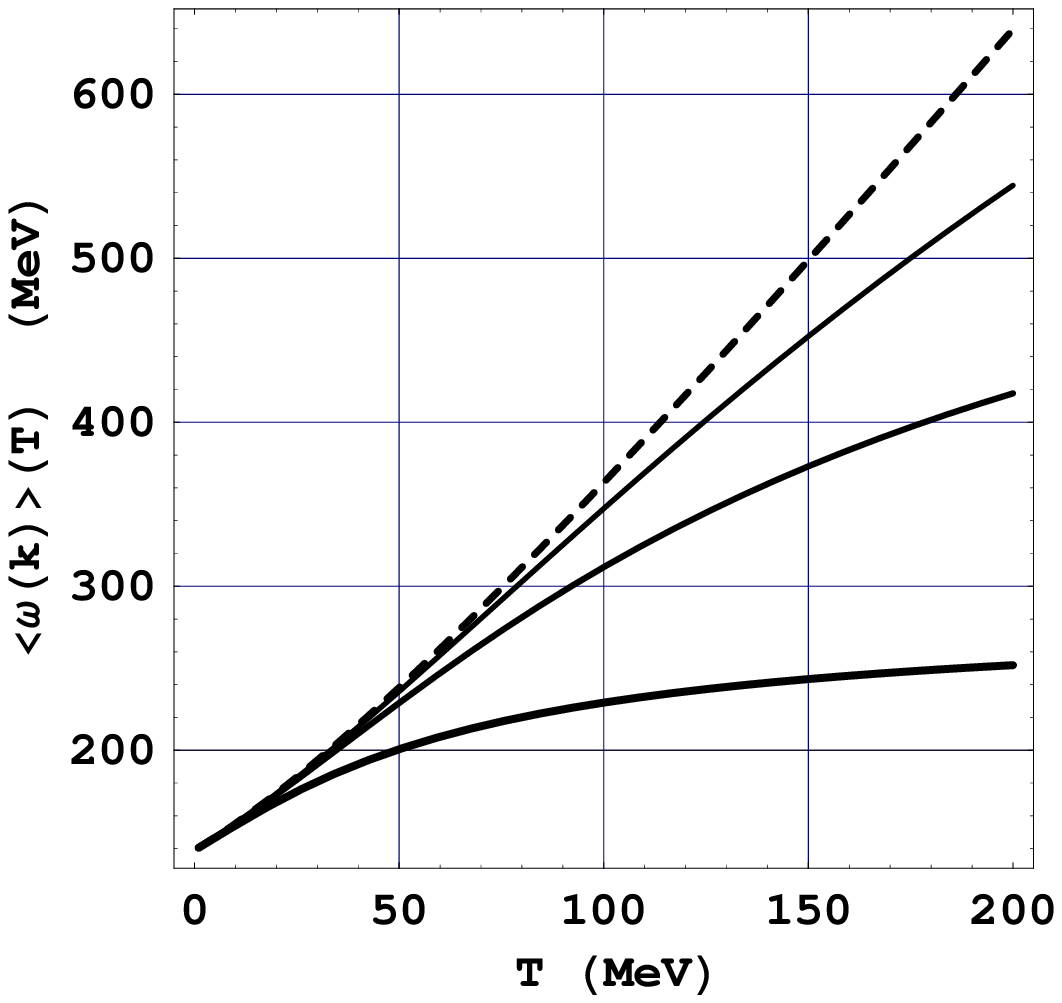}
{\footnotesize  Fig.~4.~The average single-particle energy after
5 (upper panel) and 10 (lower panel) consecutive effective scattering
for
$E_{\mathrm{tot}}$ per particle equals $200$, $400$, and $800$~MeV
(solid curves from bottom to top); and the average single-particle
energy for ideal gas of $\pi$-mesons (dashed line). }
  \end{minipage}
\end{center}
%
%
the average single-particle energy, which is equal to
$E_{\mathrm{tot}}/N$, is the same:
\[
 \frac{( \alpha/\pi )^{3/2}}{z(T)} \int \frac{d^3k}{(2\pi)^3}\,
\omega({\bs k}) \, e^{ - \frac{\omega({\bs k})}{T} - \alpha {\bs
k}^2 } =
\]
\begin{equation}
= \frac{V}{ z_{\rm id}(T_{\rm eff}) }  \int
\frac{d^3k}{(2\pi)^3}\, \omega({\bs k}) \,  e^{  -
\frac{\omega({\bs k})}{T_{\rm eff}} } \,  .
\label{205}
\end{equation}
In our illustrative example we assume that the pion energy, which
corresponds to the maximal momentum transfer, $p_{\mathrm{max}}$,
scales like the mean energy per particle:
$E_{\mathrm{tot}}/N=\sqrt{m^2_\pi+3p_{\mathrm{max}}^2}\,,$
where $m_\pi$ is the pion mass.
In Fig.~4, the results of calculations of single-particle average
energy of pions in accordance with Eqs.~(\ref{204}) (solid curves) and
(\ref{ideal-gas}) (dashed curve) are depicted as the function of
the parameter $T$ (temperature for the case of ideal gas).
The horizontal straight line in this figure corresponds to a constant
value of single-particle average energy,
$\langle \omega (\boldsymbol{k})\rangle =\mathrm{const}$.
The intersections of this straight line with a curve determine the
temperature, which corresponds to this value of the single-particle
average energy.
Therefore, fixing the average value of the energy per particle, we
come to a conclusion that the effective temperature determined by the
intersection with the dashed curve is lower than the value of the
parameter $T$, which is determined by the intersection with the solid
curve, i.e. $T_{\mathrm{eff}}\leq T$.

Such a situation completely corresponds to the relationship between
the effective temperature and parameter $T$ in non-relativistic systems
(\ref{122-5}).
Similarly to the non-relativistic case, the effective temperature is a
limit value for parameter $T$, which is reached after a sufficiently
large number of collisions $M$ occurs and/or provided that the accessible
momentum space is unconfined, i.e.
$T\rightarrow T_{\mathrm{eff}}$, if $M\rightarrow \infty $
and/or $p_{\mathrm{max}}\rightarrow \infty $.

\section{Discussion and Conclusions}
In this work, on the basis of the model proposed for the collision
of two relativistic many-particle systems (for example, nuclei),
the process of thermal equilibrium establishment has been analyzed.
This process was parametrized by the number of effective
collisions per nucleon: every effective collision brings about the
total randomization of the nucleon momentum transfer.
Note that in kinetic models, in contrast to our approach, the
processes of reaching the thermal equilibrium are parametrized by
time rather than the collision number.

The main result of the work is the explicit expression
(\ref{120prime}) for the two-source nonequilibrium distribution
function, which depends on the collision number and
the effective volume of the single-particle space of allowed
momentum transfer.
This volume is determined by the parameter $p_{\rm max}$.
There is one more essential feature of the
distribution function (\ref{120prime}):
it carries memory of the state of nuclei before collision, i.e.
the initial momentum of nucleons, $\bs k_0$, of one nucleus and
the initial momentum of nucleons, $-\bs k_0$, of another nucleus.
The immediate successful application of the two-source
distribution (\ref{120prime}) is evaluation of the single-particle
spectrum (\ref{301}).
The spectrum has two items, which can be  attributed to the first
and to the second colliding nuclei, respectively, and in
accordance with this  structure it can be named as a ``two-source
single-particle spectrum''.

In application to relativistic nucleus-nucleus collisions we extend
a statistical model where the many-particle partition function, which
describes the many-particle system just after freeze-out, is defined
as a multi-dimensional phase-space integral (see e.g.,
\cite{hagedorn-1971,bykling-1973,rischke-2001,koch-2002}).
In this approach
the single-inclusive particle
spectrum, which is created in multiparticle production process, is
attributed to the volume of $3(N-1)$ dimensional momentum
space of the unobserved $N-1$ particles
\[\omega (\bs k_1) \, \frac{d\sigma}{d^3k_1}
\propto   \left| \textsf{M} (K,k_1,\overline{k}_2,\ldots
,\overline{k}_N) \right|^2  \]
\begin{equation}
\times \,\int \prod_{i=2}^N \frac{d^3k_i}{\omega(\bs k_i)} \,
\delta^4 \! \left( K-k_1-\sum_{i=2}^N k_i \right)  \,,
\label{310}
\end{equation}%
where $k_1,\ldots ,k_N$ are the final momenta of all particles
after freeze-out with $k_i^0 \! \! =\! \omega(\bs k_i)$, $K$ is the final
total 4-momentum of the system of these particles, and
$\overline{k}_2,\ldots ,\overline{k}_N$ are the fixed values of
the momenta determined by the law of the mean when one extracts
$\left| \textsf{M} (K,k_1,k_2,\ldots ,k_N) \right|^2$ from the
integrand on the r.h.s. of (\ref{310}).
Note, $\left| \textsf{M} \right|^2$ is the modulus
squared matrix element of the multiparticle production process,
which in a simplest version of the statistical model is adopted as
a constant due to phase-space dominance.

Actually, the matrix element of the multiparticle production,
$\textsf{M}(K,k_1\ldots ,k_N)$, does not describe the interactions of
the created particles (final-state interactions \cite{rischke-2001})
before decay of the system.
At the same time, these interactions are responsible for a partial or
a full thermalization of the system.

In future, in the spirit of the present work, we plan to include in
(\ref{310}) the rescattering of the secondary particles after their
creation till the freeze-out in the same manner as it is proposed
above within the {\it maximal isotropization model}.

So, in the framework of the quasi-microcanonical concept (total energy
is fixed) of the statistical model, (\ref{310}), one cannot say how
many rescattering every particle went
through after creation and before freeze-out, moreover if there been
any rescattering at all (imaging that the system of neutrinos is
created).
Actually, without answer to this question it is not possible to shift
consideration to the canonical ensemble (temperature-fixed)
\cite{koch-2002}.
That is why, with respect just to expression (\ref{310})
and without complementary kinetic arguments, which account for
particle rescattering,
one cannot say to what extent thermalization process takes place in
the system before the particles are freezed out.

At the same time, in our approach we considered successive
collisions between
particles during evolution of the system and obtained the
nonequilibrium single-particle distribution function
(\ref{120prime}), which shows the degree of thermalization in the
system.
That is, we considered the process of thermal equilibrium
establishment as a function of the finite number of collisions per
particle, $M$, before freeze-out.
It has to be noted that the
parametrization of a nonequilibrium process by the number of
collisions per particle is especially promising in the range of
relativistic energies, because the collision number is an
invariant with respect to the change of the reference frame.

We parametrize the time axis by the number of particle collisions
$M=\langle \nu \rangle t$, where $\langle \nu \rangle$ is the mean
frequency of collisions and $t$ is the time interval.
If one considers a dependence of the distribution function
(\ref{120prime}) on the variable $M$, exactly in this sense
he can regard this distribution as nonequilibrium one
($M \propto$ time).
Increase of $M$ effectively mimics an approach to the Boltzmann
distribution (for the sake of simplicity we do not consider
special statistics in this issue).
If the number of collisions, $M$, is fixed, it means that the process
of thermalization stopped at the moment when the particles had
experienced just $M$ collisions what results in a partial
thermalization of the system.
So, the level of thermalization and isotropization depends on the
number of effective collisions, $M$, which is determined, first, by the
life time of the system or more precisely by the number of physical
collisions of every particle, $N_{\rm coll}$, and, second, by the
level of randomization of momentum transfer in every physical
collision.
As we have discussed, a degree of randomization of the particle
momentum transfer attributed to a single physical diagram (see Fig.~1)
varies from one for elastic collisions to three for inelastic
reactions, when a yield of secondary particles is large.
Note, this concerns just the hadronic processes, and we keep aside so
far all issues related to the quark gluon plasma.

So, we understand the relation of the value of the collision number
$M$ to the degree of thermalization in the system.
Let us discuss how the number of effective collisions depends on
collision energy.
For instance, we make a shift from AGS energies to SPS ones
(we fix centrality and the type of nuclei).
First, with increase of energy the number of physical hadron
collisions, $N_{\rm coll}$, decreases.
Indeed, when one considers the sequential reactions of a nucleon in
the framework of UrQMD \cite{urqmd1,urqmd2}, he finds a less number of
the physical diagrams (like that depicted in Fig.~1)
at SPS energies than at AGS energies.
Remind, the number of effective collisions is determined as:
\begin{equation}
M=\frac 13 \, \sum_{i=1}^{N_{\rm coll}}  \, d_p^{(i)} \,,
\label{defM}
\end{equation}
where $d_p^{(i)}$ is the number of random degrees of freedom gained by
every particle during $i$-th collision (see discussion after Fig.~1).
Hence, with increase of collision energy the number of items in the
sum (\ref{defM}) decreases.
On the other hand, the mean value of every item in the sum (\ref{defM})
increases because of increase of the yield of secondary particles.
However, as we showed, the creation of secondary particles can increase
the number, $M$, of effective collisions not more than by the
factor three.
So, there are two mechanisms, which determine the
number of effective collisions $M$ and hence they determine the degree
of thermalization of the nucleon system.
At the same time, these two mechanisms compete with one another with
increase of the collision energy.
For instance, in case of their balance, the dependence of $M$ on the
collision energy is absent.
It is necessary to add for completeness, that the parameter
$p_{\rm max}$, which influences also on the degree of thermalization
(see definition (\ref{12prime})), evidently increases with increase of
collision energy.

To give an explicit example we evaluate
the effective temperature of the subsystem of particles for particular
initial condition: initial momentum of particles equals zero,
$\bs k_0=0$.
This case may be regarded as a toy model, which considers the
rescattering of the particles created in nucleus-nucleus collision
(momentum of the particle just after creation is regarded as a first
momentum transfer)
or as a mathematical limit when $\bs k_0 \to 0$.
Effective temperature, $T_{\rm eff}$, is defined as a temperature of
the ideal gas of the same particles with the same mean energy per
particle.
It turns out that the temperature of
the ideal gas under this condition is always lower than the inverse
parameter $T=1/\beta$ from the nonequilibrium distribution
(\ref{120prime}) when the number of collisions, $M$, is finite.
The temperatures biggin to coincide,
$T_{\rm eff} \to T$, when collision number is big enough,
$M\to \infty$.

We hope that the distribution function derived in the present
paper will help to understand better the effect of the hadronic
rescattering in non-central relativistic collisions of heavy ions
on the azimuth anisotropy of the momentum spectra (elliptic flow)
\cite{Ma}.

A comparison of the results obtained theoretically with
experimental ones will be carried out elsewhere.

\section*{Acknowledgements:}
\vspace{-2mm}

D.A. is grateful to L.~McLerran and Yu.~Kovchegov for
useful discussions during the QM'2005 Conference which encouraged
this work.
The authors are indebted to A.~Muskeyev for providing them with UrQMD
calculations.
The authors are also grateful to F.~Becattini for discussions of the
results of the work.
D.A. was partially supported by the program \textquotedblleft
Fundamental properties of physical systems under extreme
conditions\textquotedblright\ (Section of the physics and astronomy
of the NAS of Ukraine).

\end{multicols}
\end{document}